\documentclass[aps,prd,showkeys,amssymb,preprintnumbers,
nofootinbib,twocolumn]{revtex4}

\usepackage{epsfig}
\usepackage{graphicx}
\usepackage{latexsym}
\usepackage{amsfonts}
\usepackage{url,hyperref}
\usepackage{bm}
\newcommand{\be}{\begin{equation}}
\newcommand{\ee}{\end{equation}}
\newcommand{\bea}{\begin{eqnarray}}
\newcommand{\eea}{\end{eqnarray}}
\newcommand{\bref}[1]{(\ref{#1})}

\begin{document}
\begin{figure}[t]
\vspace{-1.5cm}
\hspace{-7.5cm}
\scalebox{0.085}{\includegraphics{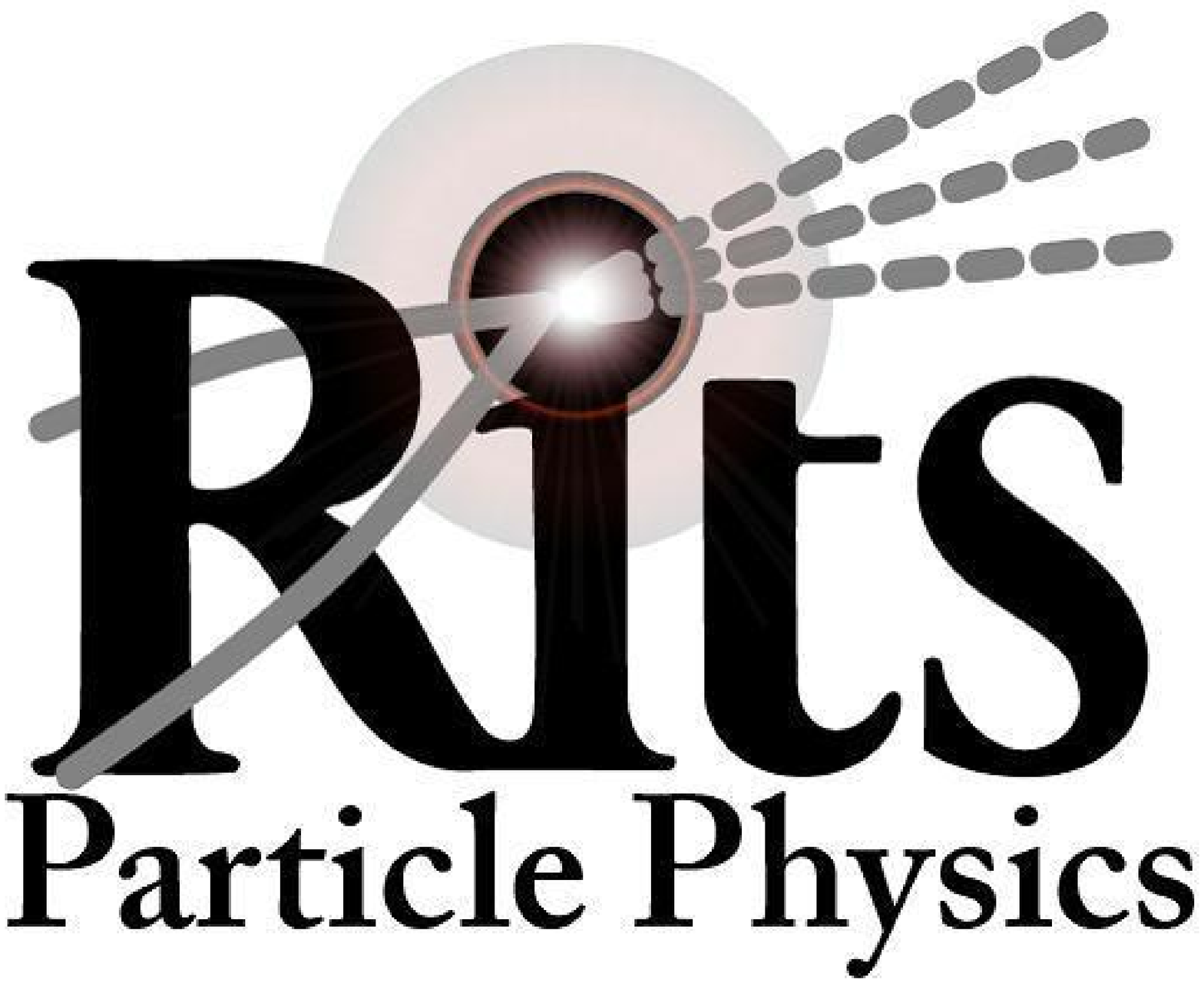}} 
\end{figure}
%%%%%%%%%%%%%%%%%%%%%%%%%%%%%%%%%%%%%%%%%%%%%%%%%%%%%%%%%
\title{Reheat Temperature and the Right-handed Neutrino Mass}
\author{Takeshi~Fukuyama}
\email[Email: ]{fukuyama@se.ritsumei.ac.jp}
\affiliation{Department of Physics, Ritsumeikan University, 
Kusatsu, Shiga 525-8577, Japan}
\author{Tatsuru~Kikuchi}\email[Email: ]{rp009979@se.ritsumei.ac.jp}
\affiliation{Department of Physics, Ritsumeikan University, 
Kusatsu, Shiga 525-8577, Japan}
\author{Wade~Naylor}
\email[Email: ]{naylor@se.ritsumei.ac.jp}
\affiliation{Department of Physics, Ritsumeikan University, 
Kusatsu, Shiga 525-8577, Japan}

\begin{abstract}
We discuss the reheating temperature in the instant preheating scenario. 
In this scenario, at the last stage of inflation, the inflaton field first decays 
into another scalar field with an enormous number density via the instant 
preheating mechanism. Subsequently, the produced scalar 
field decays into normal matter accompanied by the usual reheating mechanism. 
As an inflationary model, we identify the inflaton as a field which 
gives rise to a mass for the right-handed (s)neutrino. 
One of the interesting consequences of the instant preheating mechanism 
is the fact that the reheating temperature is proportional to the mass of 
the decayed particle, {\it the right-handed sneutrino}, $T_R \propto M_R$. 
This is very different from the ordinary perturbative reheating scenario 
in which the reheating temperature is proportional to the mass of the inflaton.
\end{abstract} 

%\pacs{04.50.+h; 98.80.Cq}
\preprint{RITS-PP-005}
\date{\today}
\maketitle

%%%%%%%%%%%%%%%%%%%%%%%%%%%%%%%%%%%%%%%%%%%%%%%%%%%%%%%%%%%%%%%%%%%%%
Inflation is a well-motivated scenario for solving many problems 
in the standard Big Bang cosmology: the flatness problem, monopole problem 
and so on \cite{Linde:2005ht}. The basic framework is constructed by using 
a single scalar field with a monomial potential. Although such a simple `toy' 
model may be attractive, it has serious difficulties from the particle 
physics point of view. This is the so called gauge hierarchy problem in 
the standard model: When we take into account the radiative corrections 
of the scalar mass, it receives quadratically divergent contributions from 
UV physics. The most promising way to solve the gauge hierarchy problem is 
to introduce supersymmetry (SUSY) \cite{Nilles:1983ge}. In models with SUSY 
it also gives a basic tool for constructing inflationary potentials in 
a rather natural way, rather than non-SUSY models, due to the enhanced 
symmetry and the fact that radiative corrections can be kept under control. 
In making an inflationary model in the non-SUSY set up we put a scalar field 
in by hand; however, in SUSY models we are fortunate to have many candidates 
for such scalar fields representing flat directions in the field configuration 
space. Indeed, there are many flat directions even in the MSSM 
\cite{Enqvist:2003gh}.

From the low energy phenomenological point of view, supersymmetric 
grand unified theory (GUT) provides an attractive framework for 
the understanding of low-energy physics. In fact, for instance, 
the anomaly cancellation between the several matter multiplets present is automatic in GUT, 
since the matter multiplets are unified into a few multiplets, 
and the experimental data supports the fact of unification of three gauge 
couplings at the GUT scale, $M_{\rm GUT} = 2 \times 10^{16}$ GeV, assuming the particle 
content of the minimal supersymmetric standard model (MSSM) 
\cite{unification1, unification2}. The right-handed neutrino, which appears
naturally in the ${\rm SO}(10)$ GUT, provides a natural explanation for 
the smallness of the neutrino masses through the see-saw mechanism 
\cite{seesaw}.

However, there is no clear connection between the reheating temperature and 
GUT scale physics, 
like for the masses of the right-handed neutrinos. Hence, we shall discuss 
the reheating process using the instant preheating mechanism and show 
that the reheating temperature is given by a mass
of the right-handed (s)neutrino.

%%%%%%%%%%%%%%%%%%%%%%%%%%%%%%%%%%%%%%%
Consider first, the following superpotential relevant for inflation
\cite{non-thermal}
\bea
W = M_I I^2 + M_{R_i} N^c_i N^c_i
+ \lambda_i I N^c_i N^c_i + Y_\nu^{ij} N^c_i L_j H_u \;,
\label{W}
\eea
where $N^c_i$ and $L_j$ are the right-handed neutrino and lepton doublet
superfields and $I$ is a complete Standard Model singlet superfield; later 
the scalar component of the singlet will be identified with the inflaton field. 

From the superpotential \bref{W}, we obtain the Lagrangian relevant for 
the preheating as follows:
\be
\label{int}
{\cal L} = - \frac{1}{2} M_I^2 I^2 
- \frac{1}{2} M_{Ri}^2 \widetilde{N^c_i}^2 
- \lambda_i^2 I^2 \widetilde{N^c_i}^2 
+ Y_\nu^{ij} \widetilde{N^c_i} L_j \widetilde{H_u} \;.
\ee
In such a model the {\it right-handed sneutrino} is coupled to the inflaton,
and after developing a VEV the right-handed neutrinos obtain their masses 
at the order of about $10^{13}$ [GeV]. Furthermore, because $Y_\nu \sim Y_u$ 
($Y_u$: up-type quark Yukawa coupling) is naturally expected in models with 
an underlying ${\rm SU}(4) \subset {\rm SO}(10)$ {\it Pati-Salam} symmetry;
hence, we can naturally expect there to be many large couplings between 
the scalar field $\widetilde{N^c_i}$ and the fermionic fields 
$L_j$ and $\widetilde{H_u}$, which are required in order to obtain a viable 
instant preheating: $I \to \widetilde{N^c_i} \to L_j \widetilde{H_u}$.

First, let us briefly consider the perturbative treatment of reheating. 
When the inflaton potential is given as above the inflaton decay rate 
is found to be
\be
\Gamma(I \to {N^c_i} {N^c_i}) 
\simeq \frac{|\lambda_i|^2}{4 \pi} M_I \;
\ee
and thus, within the perturbative treatment of reheating, 
the reheating temperature is obtained in terms of the decay rate as
\bea
T_R &=& \left(\frac{45}{2 \pi^2 g_*} \right)^{1/4}
(\Gamma M_{\rm PL})^{1/2} \ \simeq\ 0.1 \times |\lambda_i| 
\sqrt{M_I \cdot M_{\rm PL}}
\nonumber\\
&\simeq& 1.3 \times 10^{15}~{\rm GeV}~~\mbox{(for~$|\lambda_i| \sim 1$)}\;.
\label{retemp}
\eea
Here the mass of the inflaton, $M_I$, has been determined from the CMB 
anisotropy constraint,
\be
\frac{\delta \rho}{\rho}= N \sqrt{\frac{4}{3 \pi}}\frac{M_I}{M_{\rm PL}}
= 5 \times 10^{-5}
~~\Rightarrow~~
M_I \simeq 1.4  \times 10^{13} ~{\rm GeV}\,,
\ee
where we have taken the number of e-foldings to be $N\approx 60$.
In the chaotic inflation scenario, the value of the inflaton at the time 
of terminating inflation is given by 
$I_{\rm end} = M_{\rm PL}/(2 \sqrt{\pi})$, 
and the corresponding energy density is therefore
\be
\rho_{\rm end} = \frac{3}{2} V(I_{\rm end})
= \frac{3 M_I^2 M_{\rm PL}^2}{16 \pi}
= (6.5 \times 10^{15} {\rm GeV})^4\,.
\ee

Let us now return to non-perturbative reheating, i.e. {\it preheating}. 
During reheating there are in general three time scales:
\be
\begin{array}{ccc}
\left.
\begin{array}{c}
t_{\rm osc} \sim M_I^{-1} \sim 10^{-36} s
\\
t_{\rm exp} \gtrsim H_{\rm end}^{-1} \sim 10^{-35} s
\\
t_{\rm dec} \sim \Gamma_I^{-1} \sim 10^{-25} s
\end{array}
\right\}
&
\Rightarrow 
&
t_{\rm osc} \ll t_{\rm exp} \ll t_{\rm dec}
\end{array}
\ee
and so we expect several oscillations per Hubble time. Therefore, 
we would expect many oscillations of the inflaton field before it decays,
which in general leads to broad parametric resonance \cite{pre}. 
%However, as argued recently in \cite{Ahn}, instant preheating \cite{instpre} 
%may also occur if the inflaton is naturally coupled to the Higgs or another 
%scalar field.
In the usual broad parametric resonance for a hyperbolic potential, it is 
assumed that there is a succession of {\it scatterings} by the potential 
every time the field oscillates about the origin. However, there are cases 
when the field only needs to oscillate about the origin once (before rolling 
back down the potential again it decays into other particles by the standard 
reheating mechanism). This model is known as {\it instant preheating}
\cite{instpre}, which in many ways is far simpler than general parametric 
resonance theory. 
Indeed, as mentioned in \cite{instpre} under certain 
conditions one does not even need a parabolic potential, provided that 
the inflaton is coupled to another field quadratically.
Also, recently, it has been pointed out in \cite{Mazumdar} that 
the thermalization process is very slow in SUSY models
due to the presence of flat directions in the SUSY potential. 
However, in this letter, we adopt a model where the thermalization 
process occurs quickly by taking a suitable choice of parameters 
in the model.

Given any inflationary models, we would like to investigate the effects 
of preheating to generate a large decay rate for the inflaton. 
This can be achieved by using the instant preheating mechanism 
\cite{instpre}. 
%This is an off-shoot of the general broad parametric 
%resonance theory \cite{pre}, which is a generalisation of 
%the standard text book narrow parametric 
%resonance theory \cite{Landau}.
Thus, if the inflaton oscillates about the minimum of the potential 
only once it is possible to show that, see \cite{instpre,pre},
\be
n_k %= e^{-\pi \kappa^2} 
=\exp\left(-{\pi (k^2/a^2+ M_{R_i}^2) \over 
\lambda_i M_{R_i} \langle{\cal I}\rangle}\right)
\ee
and as discussed in \cite{instpre}
$M_{R_i} \langle{\cal I} \rangle$ can be replaced by $|\dot{\cal I}|$
which leads to
\be
n_k = \exp\left(-{\pi (k^2/a^2+ M_{R_{i}}^2) \over 
\lambda_i|\dot{\cal I}|}\right)  .
\ee
This can then be integrated to give the number
density for the right-handed sneutrinos, $\widetilde{N_i^c}$,
\bea\label{suppr}
n_{\widetilde{N_i^c}}&=&
{1 \over 2\pi^2} \int\limits_0^{\infty} dk\,k^2 n_k 
={({\lambda_i\dot{\cal I}})^{3/2} \over 8\pi^3}~ 
\exp\left(-{\pi  M_{R_i}^2  \over  \lambda_i|\dot{\cal I}|}\right)
\nonumber\\
&=&
{\left({M_{R_i} \lambda_i \langle{\cal I}\rangle } \right)^{3/2} \over 8\pi^3}~ 
e^{-{\pi  M_{R_i} \over \lambda_i \langle{\cal I}\rangle }}
\simeq
{M_{R_i}^3 \over 8\pi^3}\,e^{-\pi} \,.
\eea
As argued in \cite{instpre}, if the couplings are of order 
$\lambda_i\sim 1$ then there need not be an exponential suppression of 
the number density. This fact has recently been used in an interesting model 
of non-thermal leptogenesis in \cite{Ahn}.
The resultant reheating temperature from instant preheating is given by
\bea
T_R &=&
\left(\frac{30}{g_* \pi^2} \cdot 
m_{\widetilde{N_i^c}} \cdot n_{\widetilde{N_i^c}} \right)^{1/4}
\simeq 
\left(\frac{15}{4 \pi^5 g_*} \right)^{1/4} M_{R_i} e^{-\pi/4}
\nonumber\\
&\cong& 0.05 \times M_{R_i} \;.
\label{pretemp}
\eea
It should be stressed that the reheating temperature in equation 
(\ref{pretemp}), obtained from the
preheating mechanism, is proportional to the mass of the decayed particle,
{\it the right-handed sneutrino}, i.e. $T_R \propto M_R$, and does 
not depend on the inflaton mass.
This is very different from the ordinary perturbative 
reheating scenario in which the reheating temperature is proportional 
to the mass of the inflaton field, 
see Eq.~(\ref{retemp}). This characteristic of
proportionality is applicable to all the models using preheating.
A nice example is in the next to minimal supersymmetric standard model (NMSSM)
\cite{NMSSM}, where we can identify a singlet in this model as the inflaton 
\cite{FKN}. In such a case, very interestingly, the reheating temperature 
is determined by the Higgs mass: $T_R \propto m_H$.

%%%%%%%%%%%%%%%%%%%%%%%%%%%%%%%%%%%%%%%%%%%%%%%%%
To summarise, in this letter we have discussed the connection 
between the reheating temperature and the masses of the right-handed 
(s)neutrinos. The reheating process has been described as follows:
At the last stage of inflation, the inflaton field first decays 
into another scalar field with an enormous number density, via 
the instant preheating mechanism. Subsequently, the produced scalar 
field decays into normal matter accompanied by the usual reheating mechanism. 
Interestingly, the reheating temperature is proportional to the mass of 
the decayed particle, {\it the right-handed sneutrino}. We emphasise that 
this is very different from the ordinary perturbative reheating scenario 
in which the reheating temperature is proportional to the mass of 
the inflaton. 

%%%%%%%%%%%%%%%%%%%%%%%%%%%%%%%%%%%%%%%%%%%%%%%%
%\section*{Acknowledgments}
The work of T.F. is supported in part by 
the Grant-in-Aid for Scientific Research from the Ministry 
of Education, Science and Culture of Japan (\#16540269).
The work of T.K. is supported by the Research Fellowship 
of the Japan Society for the Promotion of Science (\#7336).
We thank the referee for useful comments.
%%%%%%%%%%%%%%%%%%%%%%%%%%%%%%%%%%%%%%%%%%%%%%%%%%%%%%

\end{document}